\documentclass{article}

\usepackage{arxiv}

\usepackage[utf8]{inputenc} 
\usepackage[T1]{fontenc}    
\usepackage{hyperref}       
\usepackage{url}            
\usepackage{booktabs}       
\usepackage{amsfonts}       
\usepackage{nicefrac}       
\usepackage{microtype}      
\usepackage{lipsum}
\usepackage{graphicx}
\graphicspath{ {./images/} }

\usepackage{fontawesome}
\usepackage{adjustbox}
\usepackage{xcolor}
\usepackage{subcaption}
\usepackage{amssymb}
\usepackage{amsmath}

\usepackage{pgfplots}
\usepgfplotslibrary{colorbrewer}
\pgfplotsset{compat = 1.17, cycle list/Set1-8} 
\usetikzlibrary{pgfplots.statistics, pgfplots.colorbrewer}
\usepackage{pgfplotstable}

\usepgfplotslibrary{groupplots}

\title{Trust in Software Supply Chains: Blockchain-Enabled SBOM and the AIBOM Future}

\author{
 Boming Xia\textsuperscript{1,2}, Dawen Zhang\textsuperscript{1,3}, Yue Liu\textsuperscript{1,2}, Qinghua Lu\textsuperscript{1,2}, Zhenchang Xing\textsuperscript{1,3}, and Liming Zhu\textsuperscript{1,2} \\
  \textsuperscript{1}CSIRO's Data61, \textsuperscript{2}University of New South Wales, \textsuperscript{3}Australian National University\\
\texttt{firstname.lastname@data61.csiro.au} 
}

\begin{document}
\maketitle
\begin{abstract}
The robustness of critical infrastructure systems is contingent upon the integrity and transparency of their software supply chains. 
A Software Bill of Materials (SBOM) is pivotal in this regard, offering an exhaustive inventory of components and dependencies crucial to software development. However, prevalent challenges in SBOM sharing, such as data tampering risks and vendors' reluctance to fully disclose sensitive information, significantly hinder its effective implementation. These challenges pose a notable threat to the security of critical infrastructure and systems where transparency and trust are paramount, underscoring the need for a more secure and flexible mechanism for SBOM sharing.
To bridge the gap, this study introduces a blockchain-empowered architecture for SBOM sharing, leveraging verifiable credentials to allow for selective disclosure. This strategy not only heightens security but also offers flexibility. Furthermore, this paper broadens the remit of SBOM to encompass AI systems, thereby coining the term AI Bill of Materials (AIBOM). The advent of AI and its application in critical infrastructure necessitates a nuanced understanding of AI software components, including their origins and interdependencies. 
The evaluation of our solution indicates the feasibility and flexibility of the proposed SBOM sharing mechanism, positing a solution for safeguarding (AI) software supply chains, which is essential for the resilience and reliability of modern critical infrastructure systems.
\end{abstract}


\section{Introduction}
\label{Sec:Intro}

Software supply chain security has emerged as a critical concern in both in-house and third-party software components, posing substantial threats to organizations  ~\cite{Hassija_Chamola_Gupta_Jain_Guizani_2021, Williams_2022}, particularly in critical infrastructure systems. Such vulnerabilities can result in system malfunctions or failures, as exemplified by the ChatGPT outage\footnote{\url{https://openai.com/blog/march-20-chatgpt-outage}} on March 20, 2023, caused by a bug in the open-source library \textit{redis-py}, which inadvertently exposed some users' chat history. This incident underscores the potential for far-reaching impacts in critical infrastructure, where similar vulnerabilities could lead to catastrophic outcomes.
The threat landscape is further aggravated by sophisticated cyber-attacks targeting software supply chains. 
A prominent example is the SolarWinds attack~\cite{wolff2021navigating}, where malicious actors infiltrated the company's Orion software through a trojanized update, impacting numerous government agencies and businesses. 
In addition, the notable surge in software supply chain attacks, with a 742\% annual increase from 2019 to 2022 ~\cite{sonatype:23}, underscores the pressing need for effective management of these vulnerabilities.

In response to these risks, the implementation of software bill of materials (SBOM) practices has emerged as a key strategy. An SBOM provides a detailed, machine-readable inventory of all software components, including dependency relationships, enhancing the ability to manage vulnerabilities, ensure license compliance, and track outdated components ~\cite{sbommini}. The significance of SBOMs is recognized at the highest levels, as evidenced by the U.S. government's executive order in May 2021 mandating their provision for improved security ~\cite{executive_2021}.

Nevertheless, SBOM sharing faces challenges, including SBOMs' susceptibility to tampering, the reluctance of software vendors to share complete information, and varying requirements among software procurers and users~\cite{Xia_Bi_Xing_Lu_Zhu_2023}. 
These challenges are particularly pronounced in critical infrastructure contexts where the stakes are exceptionally high.
As highlighted in a report published by the U.S. Cybersecurity and Infrastructure Security Agency (CISA)~\cite{sbomreport},
despite advancements in SBOM practices, the absence of automated and universally applicable and interoperable sharing solutions hinders effective SBOM adoption. 

Addressing these gaps, this study proposes a blockchain-based solution for SBOM sharing via verifiable credentials (VCs).
Blockchain's security, transparency, and decentralized nature make it a robust platform for cross-organizational data sharing, enhancing resistance to SBOM tampering and fostering trust~\cite{Zheng_Xie_Dai_Chen_Wang_2018} -- aspects crucial for critical infrastructure systems.
The use of VCs enables the authentication and integrity of SBOMs, facilitating fine-grained selective disclosure and zero-knowledge-proofs (ZKPs) \cite{W3C_VC, W3C_guide} of sensitive data, thus enhancing security and flexibility during SBOM sharing ~\cite{Mukta_Martens_Paik_Lu_Kanhere_2020}. The proof-of-concept experiments demonstrate the feasibility of this solution.
The main contributions include:
\begin{itemize}
    \item We undertake a comprehensive examination of typical SBOM sharing scenarios, including secure full disclosure, secure selective disclosure, and secure need-to-know disclosure (see Sections \ref{Sec:3Scenarios}). This analysis illuminates the unique challenges inherent in each scenario, thereby enhancing the understanding of the diverse landscapes and complexities of SBOM sharing.
    \item We introduce a blockchain-based architecture that integrates VCs to support progressive trust (e.g., selective disclosure, zero-knowledge proofs) for SBOM sharing. By leveraging the capabilities of blockchain and smart contracts, this solution facilitates SBOM sharing and catalyzes wider SBOM adoption.

\end{itemize}

The paper is organized as follows: Section \ref{Sec:Background} introduces SBOM, blockchain, and VCs. Section \ref{Sec:Scenario} discusses motivating scenarios for our solution. The design of our proposed architecture is detailed in Section \ref{Sec:Architecture}. Section \ref{Sec:Evaluation} presents our proof-of-concept evaluation. Related work is reviewed in Section \ref{Sec:RelatedWork}. Finally, Section \ref{Sec:DisCon} discusses the transition from SBOM to AIBOM and concludes the paper, highlighting future research directions.

\section{Background}
\label{Sec:Background}
\subsection{Software Bill of Materials (SBOM)}
An SBOM fundamentally serves as a detailed inventory of software components. As defined by the U.S. National Telecommunications and Information Administration (NTIA), an ideal SBOM should encompass eight pivotal data fields: component supplier name, component name, component version, unique identifiers (such as package url), dependency relationship, SBOM data author, and timestamp ~\cite{sbommini}.
Yet, a study by ~\cite{Xia_Bi_Xing_Lu_Zhu_2023} reveals a concerning trend: many software vendors do not adhere to these recommended fields, often withholding comprehensive SBOM information from their downstream customers. Existing prominent SBOM standards, including SPDX\footnote{\url{https://spdx.dev/}}, CycloneDX\footnote{\url{https://cyclonedx.org/}}, and SWID Tagging\footnote{\url{https://csrc.nist.gov/projects/Software-Identification-SWID}}, often mandate only a subset of the NTIA's suggested fields. For instance, CycloneDX's 1.4 Version\footnote{\url{https://cyclonedx.org/docs/1.4/json/\#metadata_authors}} omits the necessity for SBOM author(s) information.
Given this backdrop, the imperative for a secure yet adaptable SBOM sharing mechanism becomes evident.

\subsection{Blockchain}
Blockchain, characterized as a decentralized digital ledger, is fundamentally built on cryptographic protocols and distributed consensus mechanisms. These foundations ensure data immutability and resistance to tampering, making blockchain highly suitable for applications demanding trust, transparency, and security, such as financial transactions and supply chain management~\cite{Guo_Yu_2022}. The integration of smart contracts—self-executing agreements encoded on the blockchain—further enhances efficiency and autonomy by reducing reliance on intermediaries and central authorities ~\cite{Guo_Yu_2022, Zou_Lo_Kochhar_Le_Xia_Feng_Chen_Xu_2021}. 

In SBOM sharing, blockchain's intrinsic attributes directly address critical challenges, notably in data integrity and trust. Utilizing smart contracts allows for the automated management of the SBOM lifecycle, securely recording each transaction on the blockchain. This method ensures the unalterability and traceability of SBOM data, thus establishing a foundation of trust among stakeholders in the software supply chain, a factor of utmost importance in critical infrastructure systems.

Further, in the context of blockchain, Decentralized Identifiers (DIDs) \cite{brunner2020did} represent a novel identifier class enabling verifiable, self-sovereign digital identity. Unlike traditional digital identifiers managed by centralized entities, DIDs are controlled solely by their respective subjects. These identifiers, typically in URL format, link the subject to specific methods for secure and trustworthy digital interactions. This decentralization and control by the subject mark a significant shift from traditional, centrally managed digital identities, adding an extra layer of security and autonomy.
Residing on a distributed ledger like blockchain, DIDs are globally unique, persistent, and do not require a central registration authority.

\subsection{Verifiable credential (VC)}
VCs are digital credentials that allow entities to securely share and prove their information. They are designed to be cryptographically secure and tamper-proof, providing a way to authenticate individuals and their data without relying on centralized authorities or third-party verifiers. The W3C VCs Data Model ~\cite{W3C_VC} provides a common language and structure for issuing, storing, and presenting VCs, making them interoperable across different systems.

When combined with blockchain, VCs gain an added layer of security and decentralization. The blockchain ensures data integrity and immutability, while the verifiable claims within VCs provide robust means to authenticate issuers, validate information, manage revocation and expiry, and facilitate interoperability. This integration is particularly advantageous in SBOM sharing, where the authenticity of software component information is crucial. Blockchain-based VCs enable a secure and transparent mechanism to authenticate and share SBOMs, addressing key concerns such as vendor reluctance for full disclosure and the need for secure, tamper-proof documentation in software supply chains.

\section{Preliminaries and Motivating scenarios}
\label{Sec:Scenario}

\begin{figure}[]
  \centering
  \includegraphics[width=0.7\linewidth]{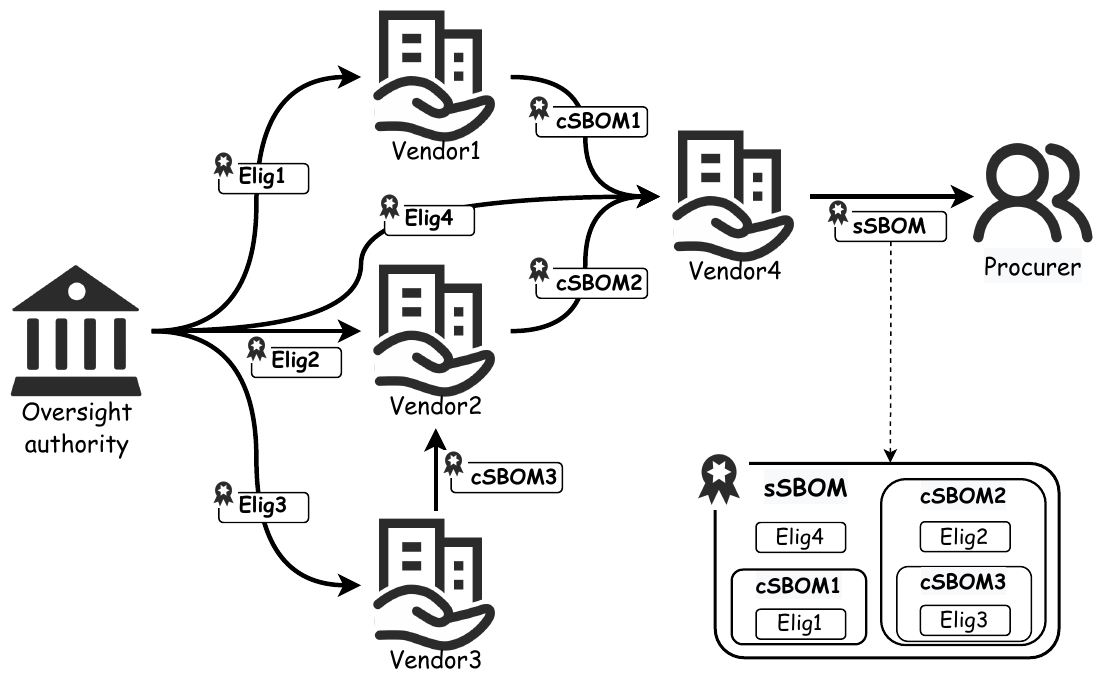}
  \caption{Trust Chain Overview (adapted from ~\cite{UN_VC:22}, Fig. 9)}
  \label{Fig:TrustChain}
\end{figure}

\subsection{Preliminaries and terminologies}
To elucidate the motivation clearly, we introduce the following preliminaries and terminologies. An overview of the Trust Chain based on VC referencing is presented in Fig. \ref{Fig:TrustChain}.

\subsubsection{VCs}
This section delineates the various VCs utilized and the establishment of a \textit{Trust Chain} through VC referencing~\cite{UN_VC:22}.

\textbf{Eligibility VCs}: Issued by oversight authorities, these VCs are granted to software vendors that exhibit adherence to security standards, industry best practices, and robust software development. They qualify vendors to issue \textit{SBOM VCs}. These VCs are referenced within \textit{SBOM VCs} using their uniform resource identifiers (URIs). It's worth noting that embedding VCs might necessitate cryptographic linking mechanisms to ensure content integrity~\cite{W3C_guide}, a topic outside this paper's purview.

\textbf{SBOM VCs}: These encapsulate SBOM metadata, linking to the actual SBOMs (via hashes or URIs) without disclosing full details. This design supports selective disclosure, allowing tailored SBOM visibility based on procurer needs or vendor policies. Given software's intricate dependencies, \textit{SBOM VCs} are further divided into:
\begin{itemize}
    \item \textbf{Component SBOM VCs (cSBOM VCs)}: For proprietary components, vendors embed their \textit{Eligibility VC}. For third-party components, existing \textit{SBOM VCs} from upstream vendors are embedded if unmodified. If absent, vendors can request them from the upstream vendor, generate their own \textit{cSBOM VCs} embedded with their \textit{Eligibility VC} after security checks, or leave them unembedded. Modifying third-party components necessitates the same steps as proprietary ones.
    
    \item \textbf{System SBOM VCs (sSBOM VCs)}: These aggregate all \textit{cSBOM VCs}, offering a comprehensive view of the software.
\end{itemize}

\subsubsection{Stakeholders}
Each stakeholder plays a critical role in ensuring the integrity and trustworthiness of the SBOM sharing chain:

\textbf{Independent Oversight (\textit{Trust Anchor})}: Oversight authorities, such as government agencies or industry-standard certification bodies, are pivotal in this solution. They establish \textbf{standardization}, ensuring consistent standards across participants. They also enforce \textbf{compliance} with legal and industry norms. Acting as a \textit{Trust Anchor}, these bodies validate and onboard new participants. While blockchain is decentralized, it still requires governance. The Oversight Authority provides this governance, underscoring that decentralization doesn't negate structured oversight. As the \textbf{issuer} of \textit{Eligibility VCs}, these authorities vet software vendors and address non-compliance, maintaining system integrity.

\textbf{Software Vendors}: Vendors, after obtaining \textit{Eligibility VCs} from oversight authorities, can self-issue \textit{SBOM VCs} for their products, embedding their eligibility within. They are the \textbf{issuer} of \textit{SBOM VCs}, the \textbf{holder} of \textit{Eligibility VCs}, and the \textbf{verifier} for both types of VCs from upstream vendors.

\textbf{Software Procurers}: These entities, which can include downstream software vendors, verify credentials during third-party software acquisitions. This verification ensures the trustworthiness of software integrated into their operations or development. They function as the \textbf{verifier} for both \textit{SBOM VCs} and \textit{Eligibility VCs}.

Note that procurers may only verify the \textit{sSBOM VCs} (and the vendor's \textit{Eligibility VC}), or, in the cases of downstream vendors, the \textit{cSBOM VCs} of the components they procure/introduce although these components can contain sub-components with \textit{cSBOM VCs}. It is the responsibility of each down-stream stakeholder to verify the VCs from the last up-stream vendor, thus forming a trust chain.

\subsubsection{Scenario overview}
\label{Sec:3Scenarios}
Multiple software vendors participate in a \textbf{permissioned} blockchain network, each responsible for their own software products. The blockchain network serves as a secure and transparent platform for sharing and verifying \textit{SBOM VCs}.

\textbf{Scenario 1: Secure Full Disclosure}.
The software vendor chooses to fully disclose the SBOM for a software product. Despite willingness for complete transparency, the challenge lies in ensuring the SBOM's integrity and authenticity against potential adversarial tampering. To counter this, secure sharing mechanisms, such as cryptographic signatures and blockchain-based records, could be leveraged to maintain the veracity and reliability of the SBOMs.

\textbf{Scenario 2: Secure Selective Disclosure}. Here, the vendor opts for partial SBOM disclosure, as the SBOM may contain sensitive information, such as proprietary algorithms or trade secrets. The disclosed parts of the SBOM align with both the vendor's policy and the procurer's requirements. This scenario employs techniques like atomic credentials, selective disclosure signatures, hashed values, and attribute-based encryption (ABE) to facilitate selective sharing while ensuring the security of the shared SBOM segments ~\cite{W3C_guide,schanzenbach2018reclaimid}. As with the first scenario, the emphasis is on secure SBOM sharing.

\textbf{Scenario 3: Secure Need-to-Know Disclosure}. This scenario is a nuanced extension of the second scenario, often employed when a vendor, adhering to strict privacy protocols, limits SBOM access. In situations like the discovery of a critical vulnerability, procurers require information about the vulnerability status in the software. The vendor, while unwilling to fully disclose the SBOM, can use ZKPs to confirm or deny the presence of the vulnerable component without revealing specific details ~\cite{W3C_VC, W3C_guide}. For example, if a critical vulnerability is known in ``\textit{component A version 2}'', but the vendor uses ``\textit{component A version 1}'', a ZKP can be constructed to exclusively demonstrate the absence of the vulnerable version. This approach maintains confidentiality, revealing only the minimal, yet critical, information.
Technically, this is executed by hashing the SBOM and associating this hash with the SBOM VC. The vendor then crafts a ZKP linked to this hash, mathematically structured to validate the specific query (presence or absence of a component) without exposing other SBOM details. The procurer, upon receiving the ZKP, cross-verifies it against the SBOM VC's hash. This process ensures the verification of the component's status while preserving the integrity and confidentiality of the complete SBOM.

In each of these scenarios, the use of blockchain and smart contracts plays an essential role. By embedding SBOM data within SBOM VCs and leveraging blockchain's capabilities, the approach guarantees both the security and flexibility of SBOM sharing, allowing vendors to control information granularity and ensuring that procurers receive dependable, verifiable SBOM data.

\section{Architecture Design}
\label{Sec:Architecture}

\begin{figure}[]
  \centering
  \includegraphics[width=0.7\linewidth]{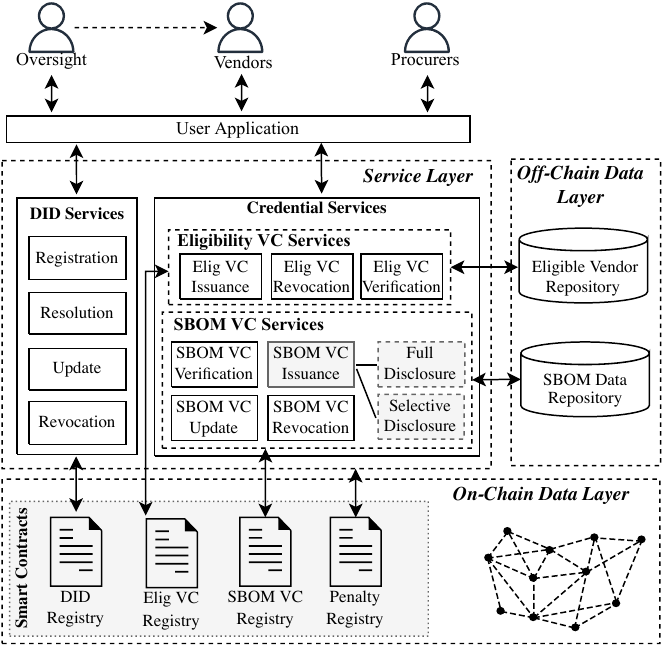}
  \caption{Architecture Overview}
  \label{Fig:Architecture}
\end{figure}

The architectural design is strategically layered (see Fig. \ref{Fig:Architecture}), balancing the decentralization, transparency, and immutability of blockchain with the efficiency and privacy of off-chain data storage. This ensures the architecture is adaptable to the diverse requirements of different SBOM sharing scenarios.
The design of each layer is crafted to address the specific needs outlined in the SBOM sharing scenarios in section \ref{Sec:3Scenarios}.

\textbf{Service Layer}: As the primary interface, this layer enables stakeholders to engage with the blockchain network and utilize its services. It integrates \textit{DID Services} and \textit{Credential Services}, which further encompass \textit{SBOM VC Services} and \textit{Eligibility VC Services}. This layer ensures authenticated stakeholder interaction, a key requirement for trustworthy SBOM sharing in all three key scenarios.

\textbf{Off-Chain Data Layer}: Operational data such as SBOM details and vendor qualifications are securely stored off the blockchain. Each software vendor independently controls their data repository, ensuring privacy and security, especially for sensitive SBOM data. The decision to utilize off-chain storage directly supports the requirements of selective and need-to-know disclosure scenarios by providing a secure environment for sensitive data, while optimizing blockchain storage efficiency and system cost-effectiveness.

\textbf{On-Chain Data Layer}: This layer contains data stored directly on the blockchain, such as VCs and DIDs, complemented by a set of smart contracts for executing specific logic. The blockchain's inherent properties ensure data immutability and transparency, bolstering data integrity and traceability, and thereby elevating system trust. In addition, this layer hosts several critical smart contracts that interact with DIDs and VCs for automated and secure system operations: 
\begin{itemize}
    \item \textit{DID Registry} Contract: Manages the registration, update, and revocation of DIDs. This contract ensures the uniqueness and immutability of each DID, providing a reliable foundation for identity verification within the network.
    \item \textit{Eligibility VC Registry} Contract: Oversees the issuance, verification, and revocation of Eligibility VCs. It ensures only compliant vendors are certified by oversight authorities. The registry also handles the revocation list, crucial for maintaining vendor eligibility integrity.
    \item \textit{SBOM VC Registry} Contract: Handles the issuance, updating, verification, and revocation of SBOM VCs. It plays a pivotal role in linking SBOM VCs to off-chain SBOM data and ensuring their validity.
    \item \textit{Penalty Registry} Contract: Enforces network compliance by managing penalties and revocations for violations.
\end{itemize}

The layered architecture is tailored to meet the functional requirements of SBOM sharing. Its adaptability allows for implementation across organizations of varying sizes, ensuring the architecture remains relevant and practical for diverse implementation contexts.

\subsection{DID Services Module}
DIDs are pivotal for managing stakeholder identities. Upon joining the network, stakeholders \textit{register} their distinct DIDs, which are archived on-chain in the \textit{DID Registry}. These DIDs authenticate subsequent actions and communications. The DID services enable DID-to-DID Document \textit{resolution}, allowing stakeholders to confirm peer identities and initiate secure interactions. Stakeholders can periodically \textit{update} their DID Documents, such as when modifying a public key or service endpoint. If a private key is compromised or a DID becomes redundant, it can be \textit{revoke}d to prevent unauthorized use and uphold network integrity.

\subsection{Credential Services: Eligibility VC Services Module}

\textit{Eligibility VC} are issued by oversight authorities (i.e., \textit{Trust Anchors}) to certify that a software vendor has met certain criteria and is therefore eligible to issue \textit{SBOM VCs}.

\subsubsection{Components}
\textbf{\textit{Eligibility VC Issuance}}: Oversight authorities \textit{issue Eligibility VCs} to software vendors who demonstrate adherence to secure software development standards and robust security practices.
These VCs are then stored on-chain, providing a transparent and immutable record of the vendor's eligibility status.

\textbf{\textit{Eligibility VC Verification}}: When a software vendor presents an \textit{Eligibility VC}, other stakeholders (like software procurers or its downstream vendors) can \textit{verify} its authenticity. This involves checking the digital signature on the VC using the public key of the issuing authority, and confirming that the VC is valid. The on-chain record ensures that the verification process is reliable and secure.

\textbf{\textit{Eligibility VC Revocation}}: If a software vendor violates the terms of eligibility, such as failing to adhere to the required standards or falsifying \textit{SBOM VCs}, the oversight authority can \textit{revoke} the vendor's \textit{Eligibility VC} by adding the VC to a revocation list.

\subsubsection{Protocol}

\begin{figure}[]
  \centering
  \includegraphics[width=0.7\linewidth]{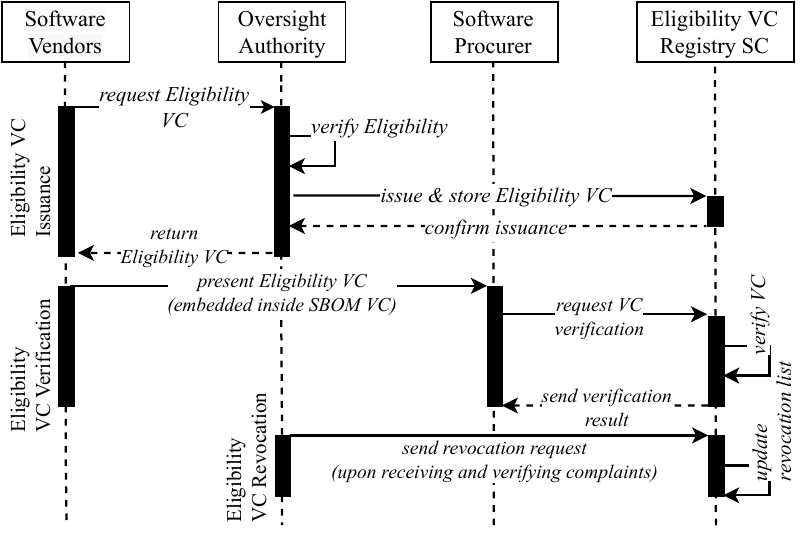}
  \caption{Protocol for Eligibility VC Services}
  \label{Fig:EligProcess}
\end{figure}

As presented in Fig. \ref{Fig:EligProcess}, the sequence of interactions commences with the software vendor applying for an \textit{Eligibility VC} from the oversight authority. The oversight authority verifies the vendor's identity and eligibility, which may encompass assessing the vendor's track record in secure software development, adherence to industry standards, and robust security practices. Once verified, the oversight authority issues an \textit{Eligibility VC} to the vendor, while the related information is stored on the blockchain and managed by the \textit{Eligibility VC registry} Smart Contract.

The vendor can then embed this \textit{Eligibility VC} within their \textit{SBOM VCs}. Software procurers, when considering a purchase, can verify the vendor's \textit{Eligibility VC} by checking its validity. This verification process involves checking the digital signature on the VC using the public key of the issuing authority, and confirming that the VC is valid and has not been revoked.

In case of any violations by the vendor, the oversight authority can invoke penalty or even revoke the \textit{Eligibility VC}. The revocation process involves adding the VC to a revocation list, which stakeholders can check when verifying an \textit{Eligibility VC}. If a VC is on the revocation list, it is no longer valid and the vendor is no longer considered eligible. These processes are managed by the \textit{Penalty Registry} and \textit{Eligibility VC registry} Smart Contracts on the blockchain, respectively.

Detailed information about the eligible vendors is stored off-chain and linked to the \textit{Eligibility VCs}. Although penalty and \textit{Eligibility VC} revocation can be invoked for various reasons, we detail this step in \textit{SBOM VC Service} protocols in Section ~\ref{Sec:SBOMProtocol} as falsifying \textit{SBOM VCs} is considered a typical reason where penalty and VC revocation are invoked.

\subsection{Credential Services: SBOM VC Services Module}
\label{Sec:SBOMVCService}
The \textit{SBOM VC Services} module is primarily managed by the \textit{SBOM VC registry} Smart Contract on-chain and plays a crucial role in overseeing the lifecycle of SBOM VCs. 

\subsubsection{Components}

\textbf{\textit{SBOM VC Issuance}}: The vendor initiates the SBOM generation process. Depending on the vendor's policies and the software procurer's requirements, the vendor can choose to fully or partially disclose the SBOM.
The \textit{SBOM VC} contains the metadata of the SBOM while excluding the actual composition details. It is then issued and stored on the blockchain, with the actual SBOM data being stored off-chain and linked to the SBOM VC.
In the case of \textbf{full disclosure}, the vendor encapsulates the SBOM metadata within the VC and signs it using their private key, without further encryption of the SBOM data.
In the case of \textbf{selective disclosure}, various techniques can be utilized to encrypt or hash the SBOM data.
For instance, to address Scenario 2 (see Section ~\ref{Sec:Scenario}), selective disclosure can be achieved through techniques such as ABE \cite{schanzenbach2018reclaimid} and hashed values \cite{W3C_guide} etc, enabling fine-grained encryption of specific (sub-)attributes. This allows for the selective disclosure of authorized attributes while maintaining the privacy of others.
Such granularity and flexibility are vital in the context of SBOMs, where component details are presented as sub-attributes of the ``Components'' (CycloneDX) or ``package'' (SPDX) attribute. A coarse-grained solution that only supports attribute-level disclosure would not suffice to address the need for partial disclosure of components.
As for Scenario 3, ZKPs can assist stakeholders in verifying the inclusion of specific data pieces within the SBOM.

\textbf{\textit{SBOM VC Verification}}: Upon issuance, any receiving party can verify the \textit{SBOM VC}. The verification process involves checking the VC's signature to ensure it was issued by a trusted vendor and has remained unaltered.
Furthermore, in the case of selective disclosure, the verification process also includes verifying the ZKPs or decrypting the VC using the decryption key provided by the vendor. This ensures that the disclosed information remains verifiable and trustworthy, while the undisclosed parts remain confidential.

\textbf{\textit{SBOM VC Update}}: As the software product evolves over time, updates to the \textit{SBOM VC} may be necessary. The vendor generates a new SBOM for the updated product and issues a new VC accordingly.

\textbf{\textit{SBOM VC Revocation}}: If the vendor determines that a previously issued \textit{SBOM VC} is no longer valid, they can revoke it. This revocation process entails adding the VC to a revocation list, which is subsequently checked during the verification process. Once a VC is included in the revocation list, it will fail the verification process and be considered invalid.

\subsubsection{Protocol}
\label{Sec:SBOMProtocol}

\begin{figure}[]
  \centering
    \includegraphics[width=0.7\linewidth]{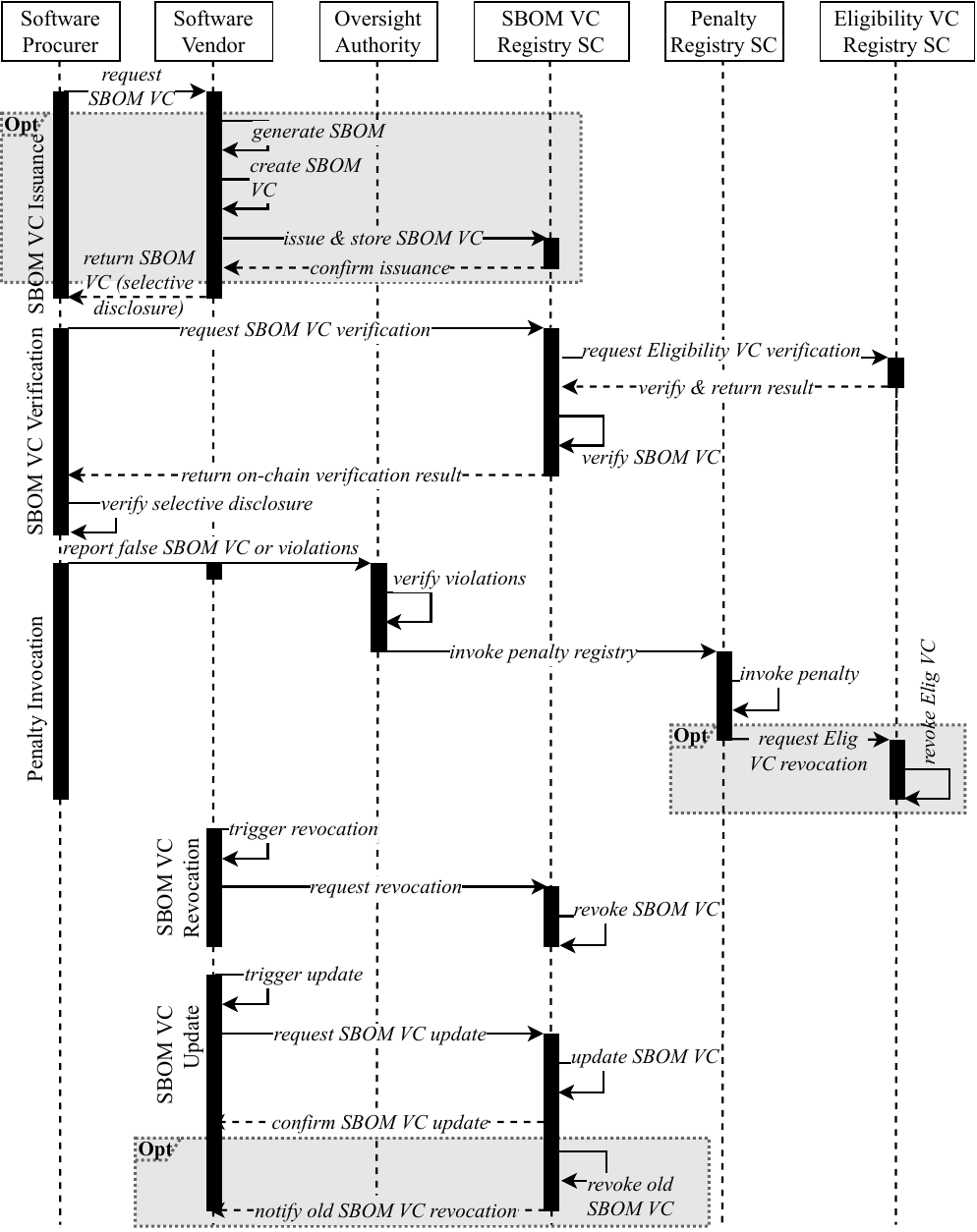}
  \caption{Protocol for SBOM VC Services}
  \label{Fig:SBOMProcess}
\end{figure}

As illustrated in Fig. \ref{Fig:SBOMProcess}, the sequence of \textit{SBOM VC Services} begins when the software procurer requests the SBOM VC from the software vendor or when the vendor proactively issues an \textit{SBOM VC} for a software product if one has not been issued yet. The vendor generates the SBOM and determines the level of information disclosure, whether it is full or selective, based on their policies and the procurer's requirements. Subsequently, the \textit{SBOM VC}, which contains the SBOM metadata without the actual composition details, is stored on the blockchain. The issuance of the requested \textit{SBOM VC} is confirmed by the \textit{SBOM VC registry} Smart Contract in coordination with the vendor. The vendor can then transmit the requested \textit{SBOM VC} to the procurer.

Upon receiving the \textit{SBOM VCs}, software procurers can verify the vendor's \textit{SBOM VC} as well as the embedded \textit{Eligibility VC} by checking their validity. The verification processes include:

\textbf{Credential verification}: This involves verifying the VC signature and checking the VC status on-chain. The VC signature is verified using the public key of the issuer (software vendor) to ensure its authenticity and integrity. The VC status is verified to ensure that the VC has not been revoked and remains valid.

\textbf{Selective disclosure verification}: In the case of selective disclosure, the software procurer may need to decrypt the VC or verify the ZKPs off-chain. This is necessary because the actual SBOM data is stored off-chain and linked to the \textit{SBOM VC}. The decryption key or the information required for ZKP verification would be provided by the vendor.

In the event of reported violations, such as falsified \textit{SBOM VCs} that are confirmed by the oversight authority, penalties can be imposed using the \textit{Penalty Registry} Smart Contract. In severe cases, the \textit{Penalty Registry} can initiate the revocation of \textit{Eligibility VCs} by invoking the \textit{Eligibility VC registry} Smart Contract.

In the event of errors or updates to the SBOM, the vendor has the ability to revoke the SBOM VC. Additionally, if the software product undergoes updates over time, the vendor would generate a new SBOM for the updated product and issue a new VC accordingly. The old \textit{SBOM VCs} can be revoked or retained for versioning purposes.

\section{Evaluation}
\label{Sec:Evaluation}
This section presents the proof-of-concept evaluations, focusing on its core components for validating the feasibility of the proposed architecture.

\subsection{Implementation}
We implemented a minimal viable prototype (MVP) for proof of concept. The prototype was built using Node.js v16.15.1, with Web3.js v1.3.6 and Solidity v0.8.2. The implementation supports the SPDX-2.2 JSON format. As described in Section \ref{Sec:Architecture}, the signatures of VCs are stored on-chain in the smart contracts, while the VC issuance, selective disclosure proof, and verification processes are assumed to occur off-chain.
Selective disclosure is facilitated through hashed values \cite{W3C_guide} in the form Merkle Hashed Tree, as depicted in \cite{Mukta_Martens_Paik_Lu_Kanhere_2020}.
The format of issued \textit{SBOM VCs} and selective disclosure proofs are also in JSON formats, containing the relevant Merkle Trees.

\subsection{Feasibility and Performance Evaluation}

\begin{figure*}
\centering
\begin{subfigure}{0.45\textwidth}
\centering
\begin{tikzpicture}
\begin{axis}[
width=\linewidth,
height=5cm,
xticklabels={\textbf{100}, \textbf{200}, \textbf{300}, \textbf{400}, \textbf{500}, \textbf{600}, \textbf{700}, \textbf{800}, \textbf{900}, \textbf{1000}},
x tick label style={rotate=45, anchor=north east, inner sep=1mm},
xlabel={\textbf{Number of Attributes}},
ylabel={\textbf{SBOM VC generation time (ms)}},
bar width=0.4cm,
grid=none,
ymin=0,
ymax=100,
xtick= {0,1,2,3,4,5,6,7,8,9,10},
xmin=-1,
xmax=10,
]
\addplot+[ybar,black,thick,error bars/.cd,y dir=both,y explicit,error bar style={line width=1pt},
error mark options={
rotate=90,
mark size=3pt,
line width=1pt
}]
plot coordinates {
(0, 28) +- (1, 4)
(1, 34) +- (5, 5)
(2, 44) +- (6, 8)
(3, 53) +- (8, 9)
(4, 59) +- (12, 9)
(5, 64) +- (12, 7)
(6, 69) +- (8, 4)
(7, 75) +- (7, 8)
(8, 83) +- (9, 8)
(9, 88) +- (9, 10)
};
\end{axis}
\end{tikzpicture}
\caption{SBOM VC generation time}
\label{Fig:Eval1GenerationTime}
\end{subfigure}
\hfill
\begin{subfigure}{0.45\textwidth}
\centering
\begin{tikzpicture}
\begin{axis}[
width=\linewidth,
height=5cm,
xticklabels={\textbf{1}, \textbf{5}, \textbf{10}, \textbf{25}, \textbf{50}, \textbf{100}, \textbf{150}, \textbf{200}, \textbf{250}, \textbf{500}},
x tick label style={rotate=45, anchor=north east, inner sep=1mm},
xlabel={\textbf{Number of Attributes}},
ylabel={\textbf{SBOM VC proof generation time (ms)}},
bar width=0.4cm,
grid=none,
ymin=0,
ymax=12,
xtick= {0,1,2,3,4,5,6,7,8,9,10},
xmin=-1,
xmax=10,
]
\addplot+[ybar,black,thick,error bars/.cd,y dir=both,y explicit,error bar style={line width=1pt},
error mark options={
rotate=90,
mark size=3pt,
line width=1pt
}]
plot coordinates {
(0, 1) +- (1, 0)
(1, 1) +- (0, 1)
(2, 1) +- (0, 1)
(3, 2) +- (1, 1)
(4, 4) +- (1, 2)
(5, 4) +- (1, 2)
(6, 4) +- (1, 2)
(7, 5) +- (1, 2)
(8, 6) +- (1, 2)
(9, 9) +- (2, 2)
};
\end{axis}
\end{tikzpicture}
\caption{SBOM VC proof generation time}
\label{Fig:Eval2ProofTime}
\end{subfigure}

\begin{subfigure}{0.45\textwidth}
    \centering
    \begin{tikzpicture}
    \begin{axis}[        width=\linewidth,        height=5cm,        xticklabels={\textbf{1}, \textbf{5}, \textbf{10}, \textbf{25}, \textbf{50}, \textbf{100}, \textbf{150}, \textbf{200}, \textbf{250}, \textbf{500}},        x tick label style={rotate=45, anchor=north east, inner sep=1mm},        xlabel={\textbf{Number of Attributes}},        ylabel={\textbf{SBOM VC proof verificationn time (ms)}},        bar width=0.4cm,        grid=none,        ymin=0,        ymax=20,        xtick= {0,1,2,3,4,5,6,7,8,9,10},        xmin=-1,         xmax=10,    ]
    \addplot+[ybar,black,thick,error bars/.cd,y dir=both,y explicit,error bar style={line width=1pt},    error mark options={        rotate=90,        mark size=3pt,        line width=1pt    }]
    plot coordinates {
        (0, 2) +- (1, 1)
        (1, 2) +- (0, 1)
        (2, 2) +- (1, 1)
        (3, 3) +- (1, 2)
        (4, 4) +- (1, 3)
        (5, 4) +- (1, 3)
        (6, 5) +- (0, 3)
        (7, 6) +- (1, 3)
        (8, 9) +- (1, 3)
        (9, 15) +- (1, 3)
    };
    \end{axis}
    \end{tikzpicture}
    \caption{SBOM VC proof verification time}
    \label{Fig:Eval3ValidationTime}
\end{subfigure}
\hfill
\begin{subfigure}{0.45\textwidth}
    \centering
    \begin{tikzpicture}
    \begin{axis}[        width=\linewidth,        height=5cm,        xticklabels={\textbf{1}, \textbf{5}, \textbf{10}, \textbf{25}, \textbf{50}, \textbf{100}},        x tick label style={rotate=45, anchor=north east, inner sep=1mm},        ylabel={\textbf{Transactions per second}},        xlabel={\textbf{Number of Threads}},        bar width=0.6cm,        grid=none,        ymin=0,        ymax=55,        xtick= {0,1,2,3,4,5,6},        xmin=-1,         xmax=6,    ]
    \addplot+[ybar,black,thick,error bars/.cd,y dir=both,y explicit,error bar style={line width=1pt},    error mark options={        rotate=90,        mark size=5pt,        line width=1pt    }]
    plot coordinates {
        (0, 12) +- (3, 2)
        (1, 27) +- (4, 5)
        (2, 39) +- (6, 8)
        (3, 38) +- (9, 7)
        (4, 37) +- (8, 5)
        (5, 40) +- (7, 6)
    };
    \end{axis}
    \end{tikzpicture}
    \caption{Throughput (SBOM VC Generation)}
    \label{Fig:Eval4Throughput}
\end{subfigure}
\caption{Performance Evaluation}
\label{Fig:Evaluation}
\end{figure*}
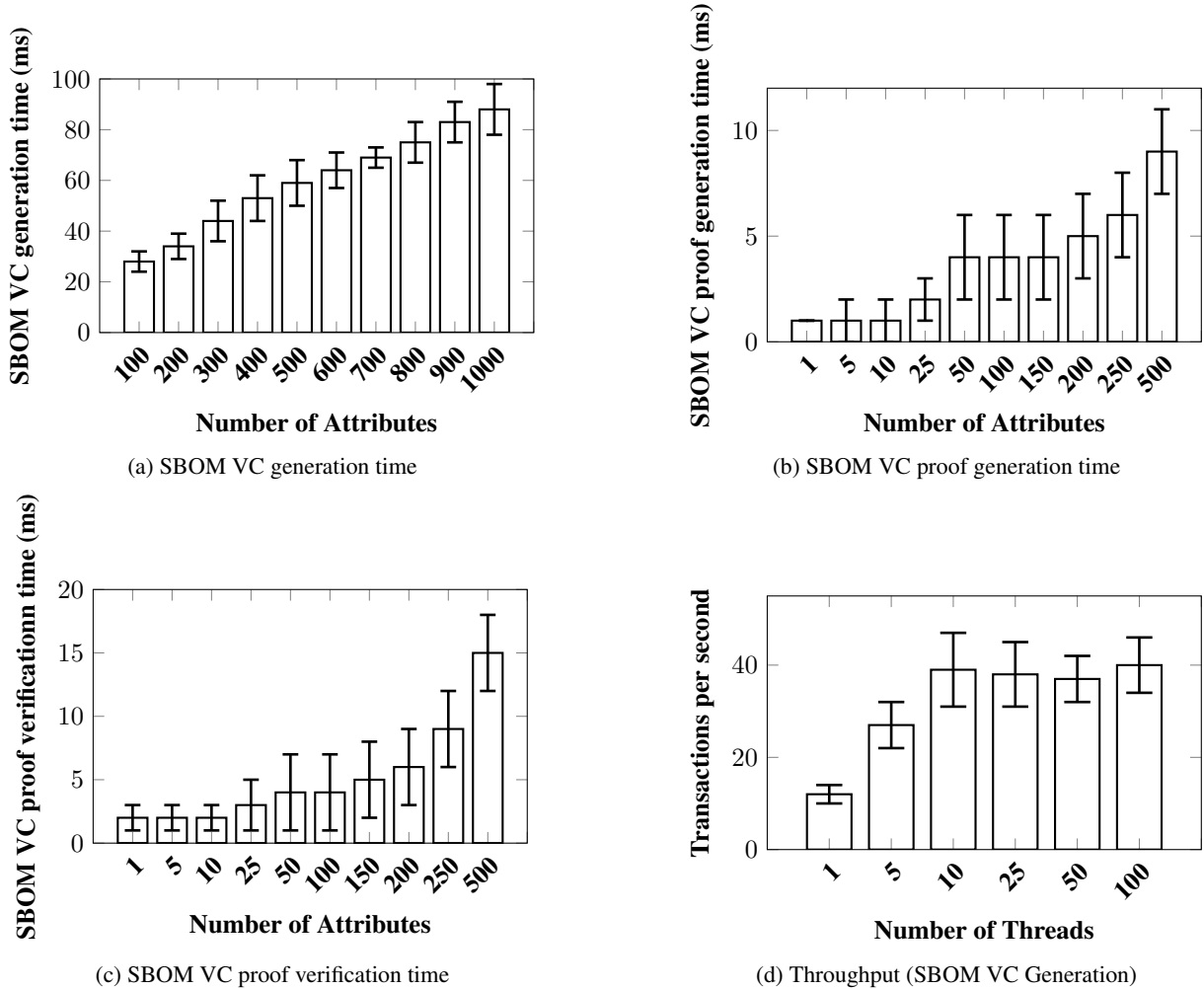

We deployed the on-chain components on a local Ganache\footnote{\url{https://github.com/trufflesuite/ganache}} blockchain network and the user application on a Google Cloud virtual machine with Ubuntu 20.04 LTS, intel Xeon E5, and 8GB RAM.
The Ganache network was configured to simulate the Ethereum Mainnet\footnote{\url{https://ethereum.org/en/developers/docs/networks/\#ethereum-mainnet}} with a Muir Glacier hardfork and an inter-block time of 12 seconds.
This setup allowed us to assess the performance of our system in a controlled environment.
To evaluate the overall responsiveness and efficiency of the deployed blockchain, we utilized Apache JMeter\footnote{\url{https://jmeter.apache.org}} to generate a load of 100 back-to-back requests. We measured the average transaction confirmation time, which yielded a value of 13.652 seconds.

Specifically, we conducted performance tests on SBOM VC generation, selective disclosure proof generation, and verification with varying numbers of corresponding attributes in the actual SBOMs. The results presented here are averaged over 20 runs. As the number of attributes increases, there is a corresponding increase in the VC generation time (see Fig. \ref{Fig:Eval1GenerationTime}). This is expected, as a larger number of attributes naturally require more computational resources to process.
Similarly, the time required for selective disclosure proof generation and verification processes increases with the number of attributes included in the proof, as shown in Fig. \ref{Fig:Eval2ProofTime} and Fig. \ref{Fig:Eval3ValidationTime}, respectively.
However, the times for selective disclosure proof generation and verification increased in a non-linear fashion. This behavior underscores the efficiency of the Merkle Tree structure, which boasts a logarithmic time complexity ($\mathcal{O}(\log n_{\text{leaf nodes}})$) for both proof and validation operations \cite{szydlo2004merkle}.
Lastly, the throughput analysis, illustrated in Fig. \ref{Fig:Eval4Throughput}, revealed the system's capability to handle between 37-40 transactions per second (TPS) across more than 10 threads.

While these evaluation results demonstrate potential scalability of our architecture, it's important to note that they are based on a controlled environment for proof-of-concept purposes and may vary in real-world applications.

\subsection{Security Analysis}
This section evaluates the architecture against the recognized web application security risks outlined in the OWASP Top Ten\footnote{\url{https://owasp.org/www-project-top-ten/}}:

\textbf{Broken Access Control}: The architecture employs permissioned blockchain to establish stringent access controls, effectively reducing risks associated with unauthorized access and broken access control.

\textbf{Cryptographic Failures}: By leveraging blockchain's inherent cryptographic functionalities for data transactions and storage, our system can mitigate cryptographic failures, ensuring secure data handling and preventing system compromises.

\textbf{Injection}: While the integration of blockchain in our architecture contributes to a reduced susceptibility to certain types of injection vulnerabilities, our current MVP does not specifically address injection vulnerabilities such as SQL Injection, which are primarily relevant for off-chain database interactions. 

\textbf{Insecure Design}: Our layered architecture incorporates secure design principles and decentralized blockchain, mitigating design flaws, including compartmentalization of responsibilities and employing immutable blockchain structures for critical data handling.

\textbf{Security Misconfiguration}:
While the current MVP implementation exhibits minimal risk of misconfiguration, we acknowledge the importance of maintaining secure configurations (e.g., regular review) in future developments.

\textbf{Vulnerable and Outdated Components}: The MVP was developed with a focus on using secure and up-to-date components to reduce the risk of vulnerabilities.

\textbf{Identification and Authentication Failures}: Leveraging DIDs and VCs, our architecture ensures robust identity management and authentication processes. These mechanisms are integrated into the blockchain framework, providing a secure and verifiable method of managing identities.

\textbf{Software and Data Integrity Failures}: Blockchain's inherent features like immutability and consensus algorithms provide strong safeguards against unauthorized alterations, thus ensuring software and data integrity.

\textbf{Security Logging and Monitoring Failures}: Blockchain's underlying distributed ledger offers an immutable record of all historical transactions, improving the system's logging and monitoring capabilities.

\textbf{Server-Side Request Forgery}: While such risks are not directly addressed in the current MVP implementation, indirect measures such as the use of VCs and the penalty mechanism for false information contribute to reducing this risk. 

\section{Related Work}
\label{Sec:RelatedWork}
The application of blockchain and VCs has been explored in various domains, but its integration into SBOM sharing presents unique challenges and opportunities.

For example, 
Arenas et al. \cite{arenas2018credenceledger} propose CredenceLedge, a permissioned blockchain for decentralized verification of academic credentials. Similarly, Mukta et al. \cite{Mukta_Martens_Paik_Lu_Kanhere_2020} introduce CredChain, a blockchain-based Self-Sovereign Identity platform that allows secure creation, sharing, and verification of credentials. They also propose a flexible selective disclosure solution using redactable signatures, emphasizing the importance of privacy in credential sharing. 

In the context of SBOM sharing, several tools and initiatives have emerged, offering different capabilities.
In assessing the current landscape of SBOM sharing tools, a selection from the CycloneDX Tool Center\footnote{\url{https://cyclonedx.org/tool-center/}} was analyzed. These tools, along with others, illustrate varied approaches in SBOM sharing, providing a comprehensive overview of how our solution addresses the needs and gaps in current SBOM sharing practices (see Table \ref{tab:sbom_comparison}).

\textbf{Sigstore's Cosign}\footnote{\url{https://docs.sigstore.dev/cosign/overview/}} enables signing and verification using a transparency log, which is an integral step for secure SBOM sharing but it does not directly support SBOM sharing. 
\textbf{SBOM.sh}\footnote{\url{https://sbom.sh/}} offers basic HTTP request-based sharing of SBOM files but lacks the multi-layered security and flexibility our architecture provides through blockchain and smart contracts.
Similarly, the \textbf{CycloneDX SBOM Exchange API}\footnote{\url{https://github.com/CycloneDX/cyclonedx-bom-exchange-api}} provide a standardized method, but they do not inherently offer the transparency and accountability provided by a shared ledger, nor do they support selective disclosure.
\textbf{RKVST}\footnote{\url{https://www.rkvst.com/share-sboms/}} approaches closer with an auditable ledger and access controls, yet it does not fully address the flexibility of complete and selective SBOM data disclosure, a gap our solution fills.

\begin{table}[]
\centering
\caption{SBOM sharing solutions}
\label{tab:sbom_comparison}
\resizebox{0.8\columnwidth}{!}{%
\begin{tabular}{|l|c|c|c|c|c|}
\hline
 & \textbf{Cosign} & \textbf{SBOM.sh} & \textbf{CycloneDX} & \textbf{RKVST} & \textbf{Ours} \\
\hline
\textbf{Method} & Sign/verify & HTTP-based & HTTP-based & \textbf{Shared Ledger} & \textbf{Blockchain} \\
\hline
\textbf{Security} & \textbf{Yes} & - & - & \textbf{Yes} & \textbf{Yes} \\
\hline
\textbf{Flexibility} & N/A & No & No & - & \textbf{Yes} \\
\hline
\end{tabular}
}
\begin{flushleft}
{\footnotesize ``-'': relevant information is not explicitly mentioned; N/A: Not applicable}
\end{flushleft}
\end{table}

Our architecture offers a comprehensive solution that not only addresses the transparency and integrity of SBOM sharing but also provides granular control over data visibility. In line with the recommendations of the CISA and CESER report \cite{sbomreport}, our solution offers the potential to promote interoperability and automation through a unified blockchain platform and smart contracts.

\section{Discussion and Conclusion}
\label{Sec:DisCon}
\subsection{From SBOM to AI Bill of Materials (AIBOM)}
With the rise of AI systems, the transition from SBOM to AIBOM becomes imperative, as underscored by advancements from leading SBOM proponents like CISA \cite{cisa_ai_sbom}, CyCloneDX \cite{cyclonedxmlbom}, and SPDX \cite{spdxaiprofile}. 
Our blockchain-based architecture is designed to accommodate this transition, leveraging VCs (linked to corresponding AIBOMs) and smart contracts to facilitate effective AIBOM sharing..

\subsubsection{Key Features of AIBOMs}
\textbf{AI-Specific Components}:
AIBOMs inherently require fields that cater to AI-specific components. This includes the integration of \textbf{models} and their associated training and testing \textbf{data}. Existing methodologies, such as datasheets for datasets \cite{gebru2021datasheets} and model cards for model reporting \cite{mitchell2019model}, can be leveraged to enhance transparency and traceability.

\textbf{Use Restrictions}:
Beyond traditional licensing scopes, AIBOMs can incorporate use restrictions that cater to ethical and societal considerations. These restrictions, which can range from prohibitions on military applications to regulations against bias, serve as a medium for communicating the \textbf{intended usage and the scope of responsible AI practices}.

\textbf{Potential Risks and Quality Issues}:
AIBOMs can also encompass potential risks and quality issues inherent to AI systems. These risks, which might arise from trade-offs in development or context-specific assessments, provide stakeholders with a comprehensive system view, facilitating informed decision-making.

\subsubsection{AIBOMs: Beyond Extended SBOMs}
The distinction between AIBOM and SBOM is not merely including additional fields. While SBOMs provide an inventory of software components, AIBOMs cater to the unique and dynamic nature of AI systems, addressing challenges intrinsic to AI's behavior and evolution.

\textbf{Dynamic Evolution and Continuous Learning}:
Traditional software, once deployed, remains static in its behavior unless updated or patched. In contrast, certain AI models, especially those like reinforcement learning algorithms, evolve as they process new data.
AIBOMs, when underpinned by the blockchain-based architecture, can chronicle this dynamic behavior, creating a verifiable ``chain of AIBOM'' that captures each evolutionary step. This not only ensures stakeholders remain apprised of AI behavioral shifts due to data influx or retraining but also facilitates rigorous auditing. Blockchain's immutable ledger can further enable features like AIBOM versioning registry, ensuring a traceable lineage, thereby enhancing accountability and transparency in AI systems.

\textbf{Implications for Liability and Regulation}:
Regulatory frameworks, such as the EU's approach to product liability \cite{stefano2023new}, are beginning to treat AI systems as products. This shift underscores the importance of tools like AIBOMs in determining liability when AI systems cause harm or make erroneous decisions.
Our architecture facilitates this by providing a detailed and immutable record of AI components and evolution, thereby assisting in liability determination and regulatory compliance. This capability is especially crucial for AI systems, where the nuances of data processing and model training can significantly impact functionality and safety.

\subsection{Conclusion}
This paper presents a blockchain-empowered solution for SBOM sharing, augmented with the use of VCs. Our architecture offers a secure, adaptable alternative to conventional SBOM sharing methods, especially crucial for critical infrastructure systems. The decentralized and immutable nature of blockchain enhances data integrity, while VCs enable selective disclosure, granting vendors precise control over SBOM data visibility.
Aligned with CISA's recommendations \cite{sbomreport}, our approach can facilitate interoperability and automation in SBOM sharing through the use of blockchain and smart contracts. Moreover, our architecture inherently supports the extension of SBOM to AIBOM, an important trajectory for future exploration in accountable AI systems.

It is important to acknowledge, however, the evaluation conducted in a controlled setting may not fully capture real-world complexities such as scalability challenges. In addition, its reliance on blockchain infrastructure may face user adoption barriers. Despite these limitations, the potential of blockchain-based solutions in this domain is evident, with similar distributed ledger based SBOM sharing solutions already being commercialized (e.g., RKVST). Future work could focus on refining our architecture to ensure its robust applicability in diverse, critical infrastructure scenarios.

\bibliographystyle{unsrt}  
\bibliography{reference}

\end{document}